\documentstyle[epsf]{mn}

\title[SZE \& Cosmology]{The SZ effect as a cosmological discriminator}
\author[J.M Diego et al.]
   { J.M  Diego$^1$, E. Mart\'\i nez-Gonz\'alez$^1$, J. L.
    Sanz$^1$, N. Benitez$^2$, J. Silk$^3$. \\
   $^1$Instituto de F\'\i sica de Cantabria, Consejo Superior de
     Investigaciones Cient\'\i ficas-Universidad de Cantabria,
     Santander, Spain\\
   $^2$Dept. of Physics \& Astronomy. The Johns Hopkins University. 
      3400 North Charles Street. Baltimore, MD 21218-2686. US\\  
   $^3$Astrophysics Dept. NAPL, Keble Road, Oxford OX1 3RH,  UK\\ }

\pagerange{\pageref{firstpage}--\pageref{lastpage}}

\begin{document}

\maketitle

\label{firstpage}
\begin{abstract}

\noindent
We show how future measurements of the SZ effect (SZE) can be used 
to constrain the cosmological parameters. We combine the SZ 
information expected from the Planck full-sky survey, $N(S)$, where no 
redshift information is included, with the $N(z)$ obtained from an 
optically-identified SZ-selected survey covering less than $1$ \% of the sky. 
We demonstrate how with a small subsample ($\approx 300$ clusters) of the 
whole SZ catalogue observed optically it is possible to drastically reduce the 
degeneracy among the cosmological parameters. We have studied the 
requirements for performing the optical follow-up and we show the feasibility 
of such a project. \\
Finally we have compared the cluster expectations for Planck with those 
expected for Newton-XMM during their lifetimes. 
It is shown that, due to its larger sky coverage, Planck 
will detect a factor $\sim 5$ times more clusters than Newton-XMM and also 
with a larger redshift coverage. 

\end{abstract}

\section{Introduction}
Clusters of galaxies have been widely used as cosmological probes. 
Their modeling can be easily understood as they are the final stage of 
the linearly evolved primordial density fluctuations 
(Press \& Schechter 1974, hereafter PS).  
As a consequence, it is possible to describe, as a function of the 
cosmological model, the distribution of clusters and their evolution, 
the {\it mass function}, which is usually used as a cosmological test 
(Bahcall \& Cen 1993, Carlberg et al. 1997, Bahcal \& Fan 1998, Girardi et al. 
1998, Rahman \& Shandarin 2000, Diego et al. 2000)  
Therefore, a detailed study of the cluster mass function will provide 
us with useful information about the underlying cosmology.\\
Following this idea, several groups have tried to constrain the 
cosmology by using the information contained in the mass function.  
They compare the observational mass function with the theoretical one 
given by the PS formalism (Carlberg et al. 1997, Bahcall \& Fan 1998) or 
with simulated ones from N-body simulations (Bahcall \& Cen 1993, 
Bode et al. 2000). 
The method has shown to be very useful but it is limitted by the quality of 
the data. Cluster masses are not very well determined for intermediate-high 
redshift clusters and even for low redshift ones the error bars are 
still significant. 
The standard methods to determine cluster masses (velocity 
dispersions, cluster richness, lensing, X-ray surface brightness deprojection) 
usually give different answers. 
It is believed that the best mass estimator is the one based on 
lensing (e.g. Wu \& Fang 1997, Allen 1998) but the number of clusters with 
masses measured with this technique  is too low to make this method a 
reliable technique to build a statistically complete mass function 
although several attempts have been made (Dahle 2000).\\
Instead of the mass function, it is possible to study the cluster population 
through other functions like the X-ray flux or luminosity functions or the 
temperature function. The advantage of these functions compared with the mass
function is that in these cases the estimation of the X-ray fluxes, 
luminosities, or temperatures of the clusters is less affected by 
systematics than the mass estimation from optical data. 
The drawback, however, is that to build these functions from the mass 
function, a relation between the mass and the X-ray luminosity, 
or flux or temperature is needed. 
These relations are known to suffer from important 
uncertainties which should be taken into account. These uncertainties have 
their origin in the intrinsic scatter of these relations as well as in the 
quality of the observational data used to build them.\\
There are three basic wavebands in which galaxy clusters are observed; 
optical (galaxies bounded to the cluster), 
X-ray (bremstrahlung emission from the very hot intra-cluster gas), 
and more recently in the mm waveband (Sunyaev-Zeldovich effect, SZE).\\  
The first clusters were observed with optical telescopes and also the 
first cluster catalogues were based on optical observations  
(Abell 1958; Abell, Corwin \& Olowin (ACO) 1989; Lumsden et al. 1992 (EDCC); 
Postman et al. 1996 (PDCS); Carlberg et al. 1997b (CNOC)). However 
the optical identification of a cluster is not a trivial task. First, 
it is not easy to define the cluster limits or cluster size. When observed 
in the optical band, a cluster appears as a group of galaxies which 
are bounded by the common gravitational potential. However, in the outer parts 
of the cluster there can be some galaxies for which it is not clear to what 
extent they are bounded to the cluster or, on the contrary, they are field 
galaxies.\\
Optical identification of clusters has other more important shortcomings. 
They specially suffer from projection effects, that is, there can be some 
galaxies in the line of sight which are not bounded by any gravitational 
force but they can appear as a bounded system because their images are 
projected onto a small circle centered on the cluster position. 
The best way to reduce this effect is by means 
of spectral identification. However, these kind of observations are time 
consuming and, when observing distant clusters, spectral identification is 
only feasible for the most luminous galaxies in the cluster.\\

\noindent
These problems are reduced when the cluster is observed in the X-ray band. 
In this band clusters appear as luminous sources due basically to the 
continuos bremstrahlung emission coming from the very hot ($T \sim 10^7$ K) 
intracluster gas. 
The same intracluster emission can be used to determine the cluster gas size. 
Moreover, projection effects are weaker in this case, thus X-ray 
surveys are very efficient in detecting clusters. 
However, with X-ray surveys it is difficult to detect clusters further 
than $z \approx 1$. There are two reasons for this. 
One lies in the fact that the X-ray cluster emissivity is proportional 
to $n_e^2$ where $n_e$ is the electron density, that is, the emissivity  
drops very fast from the center of the cluster to the border and only 
the most dense central parts of the cluster will originate a substantial X-ray 
emission. This makes very difficult the detection of distant clusters for which the 
X-ray emission is concentrated in the central parts of the cluster and 
therefore the apparent angular size will be very small. 
Moreover the X-ray flux declines as $D_L^{-2} = (D_m(1+z))^{-2}$ 
($D_m$ is the comoving distance and $D_L$ the luminosity distance). This 
selection function limits the redshift at which one cluster can be observed 
by actual X-ray detectors which are blind to the earlier stages of cluster 
formation ($z\ga 2$). \\

\noindent
In the mm waveband the situation is quite different. 
The SZE surface brightness goes as $n_e$. Therefore, 
when observing the clusters at mm wavelengths, it would be easier to 
detect the most distant ones because they will show a larger apparent 
angular size since the brightness profile drops more slowly than in the 
X-ray case. 
In X-rays, the total flux decays as $D_L^{-2}$. Meanwhile,  
the integrated SZE flux goes as $D_a^{-2}  = (D_m/(1+z))^{-2}$ 
($D_a$ being the angular diameter distance). 
$D_a$ grows more slowly than $D_L$ and even decreases after a certain 
redshift $z \approx 1$. Therefore, the SZE flux drops 
more slowly with distance (or even increases) in the case of the 
SZE than in the X-rays case. 
Another advantage is that,  as in the X-ray situation, galaxy clusters 
observed through the SZE does not suffer (almost) from projection effects. 
All these reasons make the SZE the preferable way of observing distant 
clusters.\\

\noindent
Our interest on the SZE is twofold. First, it can be considered 
as a contaminant of the cosmological signal (CMB) and therefore 
a good knowledge of this effect is required in order to 
perform an appropriate analysis of the CMB data. 
But it can also be considered as a very sensitive tool to measure the 
mass-space cluster distribution. In this paper we will concentrate our effort 
in this second aspect.\\
Planned CMB surveys will also be 
sensitive to the SZE distortion induced by galaxy clusters (MAP, Planck). 
These surveys will cover a wide area of the sky and they are expected to 
detect the SZE signature for thousands of clusters. 
Furthermore, proposed and undergoing mm experiments will measure the SZE 
for hundreds of clusters in the near future (e.g. AmiBa, Lo et al. 2000; 
LMT-BOLOCAM, L\'opez-Cruz \& Gazta\~{n}aga 2000; 
CBI, Udomprasert et al. 2000; AMI, Kneissl et al. 2001;  
or the interferometer proposed in Holder et al. 2000)
The cosmological possibilities of these  
new data sets are very relevant as they will have a statistically large  
number of clusters and they will improve the redshift coverage significantly.
For a realistic prediction of the power of future SZE surveys we 
should consider the detector characteristics of one of these planned 
experiments. 
In this work we will focus our attention on the expected SZE detections 
for the Planck mission and we will study the possibilities of these data 
to probe the cosmological model. 
Planck will observe the whole sky at 9 mm 
frequencies (including those where the SZE is more relevant) and with angular 
resolutions ranging from 5 arcmin to 33 arcmin FWHM (see fig. \ref{fig_fx} 
below). The observation of the SZE at different frequencies will 
make easier the task of identifying clusters because of the peculiar 
spectral behaviour of the SZE signal which can be very 
well recognized with the nine Planck frequencies. 
As can be seen from figure (\ref{fig_fx}), the best channels 
are the ones at 100 GHz ($x = 1.76$),  143 GHz ($x = 2.5$) and 
353 GHz ($x = 6.2$) together with the channel at 217 GHz ($x = 3.8$) 
where the thermal SZE vanishes. 
A cross correlation of these channels, including the knowledge of 
the spectral shape factor, will allow to discriminate among the SZE, 
foregrounds and CMB.\\

\noindent
As we will show in the next sections, the cluster population at high redshift 
is much more sensitive to the cosmological parameters than the cluster 
population in the local Universe ($z \la 0.5$). 
Thus, the information at high redshift is crucial 
to determine the underlying cosmology and the SZE can be the tool to 
get that information. \\

\noindent
The structure of this paper is the following; 
in section \ref{section_SZE} we recall some of the basics of the SZE and some 
useful definitions which will be of interest in subsequent sections. 
In section \ref{section_probe}, we show how through the SZE we can 
investigate the cluster population and evolution. 
We also show in section \ref{section_Optical} how  future SZE surveys should be 
complemented with optical observations of a small subsample of 
SZE-selected clusters in order to provide redshift 
information of those clusters; in this way, it is possible to reduce 
the degeneracy in the cosmological models describing the data. 
We apply this idea in section 
\ref{section_Constraining} to simulated Planck SZ data. In that section  
we obtain some interesting results about the possibilities of those future 
data. Finally we discuss our results and we conclude in section
\ref{section_Discussion}.

\section{The Sunyaev-Zel`dovich effect }\label{section_SZE}
Since Sunyaev \& Zel'dovich predicted that clusters would distort 
the spectrum of CMB photons when they cross a galaxy cluster 
(Sunyaev \& Zel'dovich 1972), 
several detections of this effect have been made. 
At present, the number of clusters with measured SZE is small because 
they are limitted by the detector sensitivity (a typical SZE signal 
is of the order of $10^{-4}$ in $\Delta T/T$ with $\bar{T} \approx 2.73$ K). 
However, the sensitivity in the detectors is getting better and better and 
in a few years the number of detected clusters through the SZE will 
make this effect an important tool to explore clusters at high redshift.\\

\begin{figure}
   \begin{center}
   \epsfxsize=8.cm 
   \begin{minipage}{\epsfxsize}\epsffile{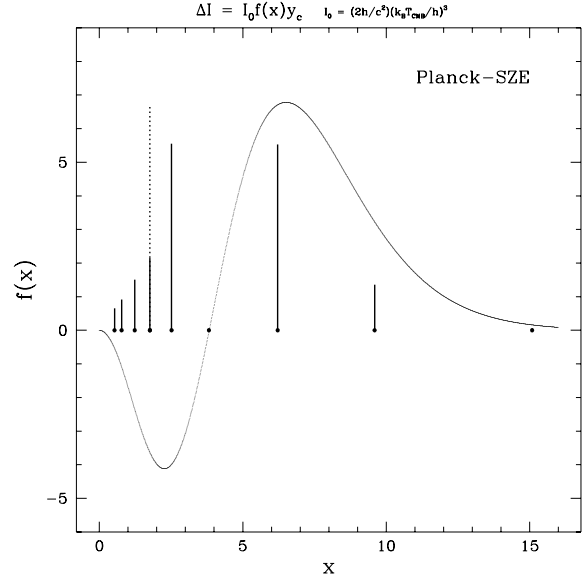}\end{minipage}
   \caption{\label{fig_fx}
            Spectral shape of the thermal SZE. 
	    Planck channels are between $30$ GHz 
            and $800$ GHz (black dots), including also the $217$ GHz channel 
	    where the thermal SZE vanishes. The best channels to detect the 
	    SZE decrement/increment will be those of the HFI instrument 
	    at 100 GHz ($x = 1.7$), 143 GHz ($x = 2.5$) 
	    and 353 GHz ($x = 6.2$) respectively together with the one at 217 
	    GHz ($x=3.8$). At 100 GHz ($x = 1.7$) there 
	    are two channels: the LFI (solid line) and the HFI (dotted line).}
   \end{center}
\end{figure}

\noindent
When CMB photons cross a cluster of galaxies, the spectrum 
suffers a distortion due to inverse Compton scattering. 
The net distortion in a given direction can be quantified by the cluster 
Comptonization parameter, $y_c$, which is defined as 
\begin{equation}
y_c = \frac{\sigma_Tk_B}{m_ec^2}\int T n_e(l)dl .
\label{y_c2}
\end{equation}
The integral is performed along the line 
of sight through the cluster. $T$ and $n_e$ are the intracluster 
electron temperature and density respectively;  
$\sigma_T$ is the Thomson cross section which is 
the appropriate cross section in this energy regime and $k_B$ is the 
Boltzmann constant.  \\
The flux and the temperature distortion are 
more widely used than the Compton parameter because these are 
the quantities which are directly determined in any 
experiment when observing clusters at mm  frequencies.
The distortion in the background temperature is given by: 
\begin{equation}
 \frac{\delta T}{T} = g(x)y_c  \ ,
\end{equation}
where $x=h\nu /kT_{CMB}$ is the frequency in dimensionless units and 
$g(x) = xcoth(x/2) - 4$ is the spectral shape factor. 
There is a second kind of distortion due to the bulk velocity of the cluster. 
It is known as the kinetic SZE. However, this contribution is much 
smaller than the thermal contribution and we will not consider that in 
this work.\\
As can be seen from the previous equation, there is no redshift dependence 
on the temperature distortion, i.e, the same cluster will induce the same 
distortion in the CMB temperature, independently of the cluster distance 
(except for relativistic corrections). 
The only redshift dependence of the total SZE flux is due to the fact 
that the apparent size of the cluster changes with redshift.\\
The total SZ flux is given by the integral :
\begin{equation}
 S_{total}(\nu) = \int_{cluster} \Delta I(\nu,\vec{\theta}) d\Omega ,
 \label{S_total}
\end{equation} 
where the integral is performed over the solid angle subtended by the cluster. 
$\Delta I(\nu,\vec{\theta})$ 
is the change in intensity induced by the SZE and is given by  
$\Delta I(\nu,\vec{\theta}) = I_0f(x)y_c 
$ where $I_0 = (2h/c^2)(k_B T_{CMB}/h)^3$ and 
\begin{equation}
f(x) =  \frac{x^4e^x}{(e^x-1)^2} \left [ \frac{x(e^x+1)}{(e^x-1)} - 4\right ] ,
\label{fx}
\end{equation}
is the spectral shape (see fig.\ref{fig_fx}).\\
In fig. \ref{fig_fx} we show the spectral shape of the SZE $f(x)$. In 
the same plot the 9 frequency channels from Planck are shown as black dots. 
The amplitude of the lines are proportional to the SNR  
per resolution element in each channel (assuming that the clusters 
are unresolved). 
The SNR  depends on the spectral shape factor $f(x)$ since the signal is 
proportional to $f(x)$ (see eq. \ref{S_total2} below) and 
the sensitivity per resolution element ($\sigma$); 
$SNR = S/N \propto f(x)/\sigma$ which is proportional to the amplitudes 
represented in fig. \ref{fig_fx}.\\

\noindent
Eq. (\ref{S_total}) can be easily integrated 
assuming that the cluster is isothermal. Then the integral can be reduced to 
$\propto \int d\Omega \int n_e(\vec{\theta})d\ell$ which can also be 
transformed into 
$D_a^{-2}(z)\int dV n_e(\vec{\theta}) =  D_a^{-2}(z) \frac{M_{gas}}{m_p} 
=  D_a^{-2}(z) \frac{f_b}{m_p}M$. $M$, $f_b$ and $m_p$ are the total mass, 
baryon fraction and the proton mass respectively.
In this approach, we have assumed that the gas is only composed of 
ionized Hydrogen. Finally we get:
\begin{equation}
 S_{SZE} = \frac{3.781 f(x) f_b T M_{15}}{D_a^2(z)} .
 \label{S_total2}
\end{equation}
In the previous expression $T$ is given in Kelvin, $M_{15}$  
in $10^{15} M_{\odot}$ and $D_a(z)$ in Mpc. 
In these units, the flux is given in mJy 
(1 mJy = $1.0\times 10^{-26}$ erg s$^{-1}$ Hz$^{-1}$ cm$^{-2}$). If we 
now introduce the $h$ dependency in $M_{15}$ ($10^{15} h^{-1} M_{\odot}$), 
$D_a(z)$ ($h^{-1}$ Mpc) and $f_b \approx 0.06 h^{-1}$ 
then the flux is given also in mJy units.\\

\noindent
Eq. (\ref{S_total2}) tell us that the total flux is given basically by three 
terms: temperature, mass and angular-diameter distance to the cluster.
As a difference with the X-ray flux, where in the integral over the 
cluster volume one should consider $n_e^2$ instead of just $n_e$, the total 
SZE flux does not depend on the cluster profile. 
Therefore, for an isothermal cluster, no assumption about the 
electron density profile is needed when computing the total SZE flux. 
This aspect is very important since for the Planck resolution, 
only a small percentage of the clusters ($\approx$ 1 \%) 
will be resolved and as a first approximation we can consider that 
all the fluxes can be computed from eq. (\ref{S_total2}). 
This fact will simplify significantly our calculations 
and simultaneously will reduce the uncertainty due to the lack of a precise 
knowledge of the cluster density profile.\\
The second important consequence that we can remark from the previous 
equation is about the angular-diameter distance dependence of the total flux. 
In X-rays, the total flux depends on $D_L^{-2}(z)$. If we compute the 
ratio $D_a^2(z)/D_L^2(z)$ it goes like $(1+z)^{-4}$, that is,
the $S_{SZE}$ drops a factor $(1+z)^{-4}$ slower than the $S_X$. Thus for 
high redshift clusters it is evident the advantage of 
using mm surveys (SZE) as compared with the X-ray ones.\\

\noindent
If the cluster is unresolved by the antenna, then the cluster total flux 
(eq. \ref{S_total2}) can be considered contained in the resolution 
element of our antenna. 
If on the contrary, the cluster is resolved, then the previous 
approximation is not adequate anymore and we should integrate the surface 
brightness over the cluster solid angle. \\
The gas density profile can be well fitted by a $\beta$-model.
\begin{equation}
 n_e(r) = n_o(1 + (r/r_c)^2)^{-3\beta /2} ,
 \label{n_e}
\end{equation}
where $n_o$ is the central electron density 
($n_o \approx 2.0 \times 10^{-3}(1+z)^3 h^{1/2} e^{-}/cm^3$) and 
$r_c$ is the core radius of the cluster. The observed values of $\beta$, 
obtained from X-ray surface brightness profiles, typically range from 
0.5 to 0.7 (Jones \& Forman 1984, Markevitch et al. 1998). 
\section{SZE as a probe of the cosmological model}\label{section_probe}
As we mentioned in the introduction, an estimate of the cluster mass 
function can be used to constrain the cosmological parameters. However 
such estimate is affected by the systematics in the mass estimators.
It is possible to use other functions to explore the cluster population 
like for instance the temperature function $d^2N/dTdz$ 
(Henry \& Arnaud 1991, Eke et al. 1998, Viana \& Liddle 1999, Donahue \& Voit 1999, 
Blanchard et al. 2000, Henry 2000). The connection between 
the mass function and the temperature function is the $T-M$ relation, 
\begin{equation}
\frac{d^2N}{dTdz} = \frac{d^2N}{dMdz}\frac{dM}{dT}
\label{dNT}
\end{equation}
Temperature estimates are more reliable than mass estimates. Therefore the 
temperature function should be less affected by the systematics than the 
mass function. However, the scatter in the $T-M$ relation introduces new 
uncertainties which should be taken into account. Similar problems have the 
X-ray luminosity function and the X-ray flux function which have been widely 
used in the literature to constrain the cosmological parameters 
(Mathiesen \& Evrard 1998, Borgani et al. 1999, Diego et al. 2000). 
In those cases new uncertainties appear due to the scatter in the $L_x-T$ relation. 
In a previous work (Diego et al. 2000), we have considered these ideas and  
applied them to constrain the cosmological parameters by fitting the model to 
different data sets: mass function, temperature function, X-ray luminosity and 
flux functions. That work was innovative in the sense 
that we considered all the previous data sets simultaneously in our fit 
(and not just one as usual) and we looked for the best fitting model to all 
the data sets considered. 
Also important is to remark that in that work we considered the cluster 
scaling relations as free-parameter relations. 
These two points are very relevant because when fitting cluster data sets  
it is important to check that the best fitting models agree well 
with other data sets. Otherwise, we should reexamine the assumptions made 
in the model. One of these assumptions is, for instance, the cluster scaling 
relations $T-M$ or $L_x-T$. These relations are commonly  
fixed a priori. However they are known to suffer from important 
scatter as we have shown in that work. In fact, we found that not all the 
scaling relations considered previously in other works were appropriate to 
describe conveniently several cluster data sets in a simultaneous fit.
By fitting our free-parameter model to all the data sets, we obtained, 
not only the cosmological parameters, but also the best scaling parameters. 
In that work we found as our best fitting parameters the ones given 
in table \ref{table}.
\begin{table}
\large
\caption{Best $\Lambda$CDM (flat)  and OCDM ($\Lambda = 0$) models. 
Both models are undistinguishable in the redshift regime of the data 
considered in Diego et al. (2000). }
\label{table}  
\normalsize
\begin{tabular}{lcccccc}
\hline
\hline
Model        & $\sigma_8$ & $\Gamma$ & $\Omega_m$ & $T_0 (10^8 h^{\alpha} K)$ & $\alpha$ & $\psi$ \\
\hline
\hline
$\Lambda$CDM &    0.8     &    0.2   &     0.3      &       1.1               &   0.75   &   1.0  \\
\hline
OCDM         &    0.8     &    0.2   &     0.3      &       1.1               &   0.8    &   1.0  \\
\hline
\end{tabular}
\label{table_best}
\end{table}
The first 3 columns correspond to the cosmological parameters and the 
others are for the $T-M$ relation;
\begin{equation}
  T_{gas} = T_0 M_{15}^{\alpha}(1 + z)^{\psi} ,
  \label{Tx}
\end{equation}
where $M_{15}$ is the cluster mass in $10^{15} h^{-1} M_{\odot}$ and $T_0$, 
$\alpha$, and $\psi$ were free parameters.
We found that the two models in table \ref{table} were undistinguishable 
when they were compared with the data. 
In fact, both models agree well with the data considered and there was no way to 
determine which one was the best. 
In order to distinguish them, more and better quality data are needed. 
Undergoing X-ray experiments (CHANDRA, Newton-XMM) will help to do that 
as well as current and proposed optical surveys (e.g. Gladders et al. 2000). 
The models which were undistinguishable in the low redshift interval will show different 
behaviour at higher redshift. However, as we mentioned in the introduction, the 
waveband in which distant clusters will be detected the most, is not the optical 
nor the X-ray band, but the mm band. The usefulness of the SZE as a tool 
to measure cosmological parameters motivated many works in the past 
(e.g. Silk \& White 1978, Markevitch et al. 1994, Bartlett \& Silk 1994, 
Barbosa et al. 1996, Aghanim et al. 1997) as well as in the present 
time (Holder et al. 2000, Grego et al. 2000, Mason et al. 2001, this work).\\

\begin{figure}
   \begin{center}
   \epsfxsize=8.cm 
   \begin{minipage}{\epsfxsize}\epsffile{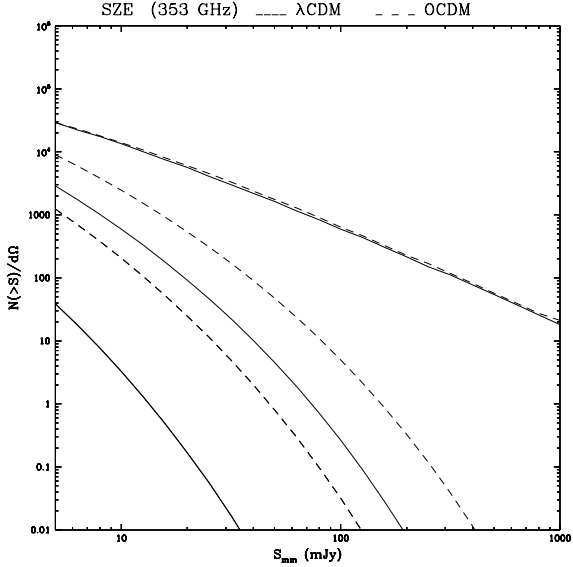}\end{minipage}
   \caption{\label{fig_NS_z_A}
            The SZE integrated number counts of the cluster 
            population predicted at mm wavelengths (353 GHz) for the best 
            flat (solid) and open (dotted) models in table 1. 
	    In the plot three redshift shells are 
            represented: top $z < 1$, middle $z \in [1,2]$ and 
            bottom $z > 2$.}
   \end{center}
\end{figure}

\begin{figure}
   \begin{center}
   \epsfxsize=8.cm 
   \begin{minipage}{\epsfxsize}\epsffile{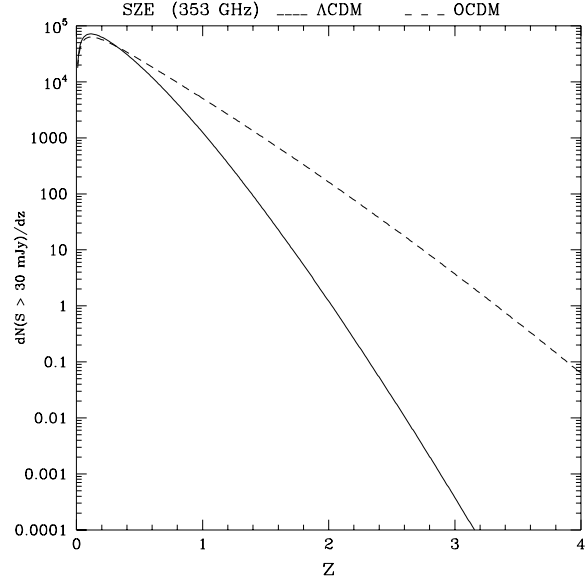}\end{minipage}
   \caption{\label{fig_dNS_z_log}
            The SZE differential number counts of the cluster 
            population predicted for the Planck mission 
	    at mm wavelengths (353 GHz) for the best 
            flat (solid) and open (dotted) models in table \ref{table} 
            as a function of redshift. 
	    }
   \end{center}
\end{figure}

\noindent
Using SZ data it is possible to estimate the cosmological parameters by looking at some 
SZ derived function related with the mass function. 
We just need a measurement of a cluster quantity related with its mass. 
As we have shown in eq. (\ref{S_total2}), the total SZ flux can 
be such a quantity. 
From that equation it is possible to build a mm flux function 
($dN/dS_{mm}$) from a mass function ($dN/dM$).  
An interesting alternative to this approach can be found in Xue \& Wu (2001). 
In that paper the authors have {\it connected} the mm flux function with 
the observed X-ray luminosity function through the $L_x-T$ and $T-M$ relations. 
If we consider a  future mm experiment (like Planck) where thousands 
of clusters are expected to be detected through the SZE, probably a fit to the 
mm flux function could be able to distinguish the models in table \ref{table}.
In fig. \ref{fig_NS_z_A} we show the expected number of clusters 
with total flux above a given flux for the two previous models.  
These are integrated fluxes, i.e we assume that the clusters are not 
resolved and that their total flux falls into the antenna beam size. 
This assumption is appropriate when the antenna size is above several arcmin  
as it is the case for the Planck satellite where its best resolution is 
FWHM = $5$ arcmin.\\
In fig. \ref{fig_NS_z_A} we did not show the whole ($0 < z < \infty $) 
integrated $N(>S)$ curve since it looks quite similar to the two curves 
on the top of the figure and they still look undistinguishable. 
To compare them in a more realistic way we have performed a 
Kolgomorov-Smirnov test where we have compared the $N(>S)$ curve for   
a realization (over 2/3 of the sky) of the $\Lambda$CDM model with the 
mean expected values of both OCDM and $\Lambda$CDM models. 
The KS test was unable to distinguish between the models.
As can be seen from the figure, both models 
predict the same number of clusters in the $z < 1$ redshift interval (top). 
That situation was similar in Diego et al. (2000) where most of the cluster 
data  were at low $z$ and it was not possible to discriminate between both 
models. \\
But the situation changes at redshifts $z > 1$. In the redshift bin 
$z \in (1,2)$ (middle) the differences between both models are significant. 
These differences are increased when we compare the models in the redshift bin 
$z > 2$ (bottom) where in the OCDM model two orders 
of magnitude more clusters are expected than in 
the $\Lambda$CDM case above $S = 30 $ mJy. 
This flux ($S = 30 $ mJy at 353 GHz) is expected to be the flux 
limit of Planck (see Appendix A where we estimate that limit based on 
MEM residuals). 
However, this limiting flux will depend on the method used to identify 
such clusters in the maps. 
The final method to be used with Planck is still under development. 
MEM methods work very well (Hobson et al. 1998, Hobson et al. 1999) 
being at present the preferred one. \\
In fig. \ref{fig_dNS_z_log} we show the number of clusters with fluxes above that 
limiting flux of Planck as a function of redshift. 
Again the differences are not significant below $z \approx 1$ but they are 
quite relevant above that redshift. By looking at 
figs. (\ref{fig_NS_z_A}) (\ref{fig_dNS_z_log}), we see that the differences 
in the cosmological models are more evident at high redshift. In order to 
discriminate among the cosmological models one should consider not only 
the cluster population at low redshift (normalization) but the cluster 
population at high redshift (evolution) as well. A recent application of this 
idea can be found in Fan \& Chiueh (2000) where the authors study the 
function $r = N(z<0.5)/N(z>1)$ as a function of the limiting flux. 
That work suggests that a limitted knowledge of the redshift of the cluster 
would allow to constrain the cosmological parameters.  \\

\noindent
Both fig. \ref{fig_NS_z_A} and fig. \ref{fig_dNS_z_log} suggest 
that with future SZE data it will be possible to go further on the 
determination of the cosmological parameters. 
From all these plots, it is evident that Planck (together with redshift 
information for a small subsample, see next section) would be able to 
discriminate between the models which previously were undistinguishable 
when they were compared with present X-ray and optical data. Furthermore, the 
accuracy in the free parameters obtained in previous works will 
be increased with these new data. \\

\subsection{Montecarlo simulations: Planck vs Newton-XMM}
The previous discussion was based on theoretical expectations of the 
mean number of SZE detections expected at different redshifts. We want to go 
further by computing Montecarlo (Press-Schechter) realizations of the 
expected SZE on a specific patch of the sky. 
In order to compare the mm and X-ray bands we will also compare the 
expected SZ detections in this area for Planck with 
those based on X-ray expectations for Newton-XMM in the same sky patch.\\

\noindent
The simulations were done over a patch of the sky of 
$12.8^{\circ}\times 12.8^{\circ}$ and with a pixel 
size of $1.5'\times 1.5'$ filtered with a FWHM of $5$ arcmin and at the 
frequency of 353 GHz, following the characteristics of the 
353 GHz channel of Planck.\\
The parameters of this simulation corresponds to the $\Lambda$CDM model 
of table 1.\\

\begin{figure}
   \begin{center}
   \epsfxsize=9.cm 
   \begin{minipage}{\epsfxsize}\epsffile{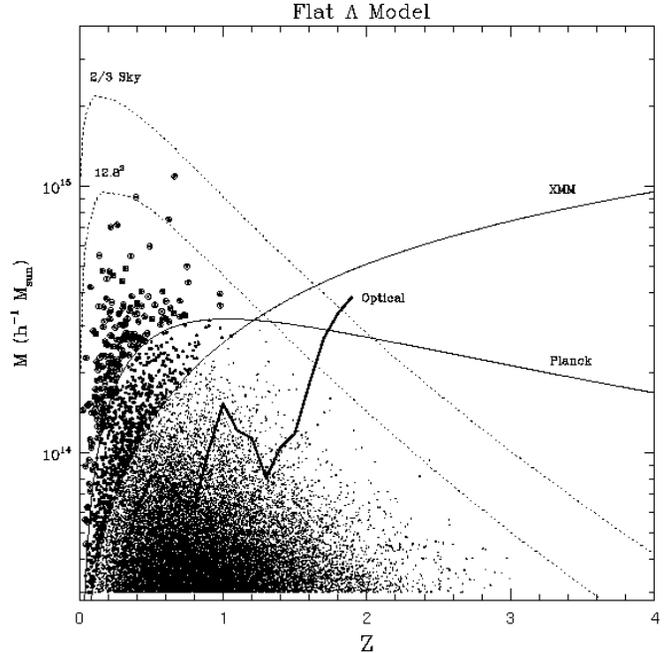}\end{minipage}
   \caption{\label{fig_Planck_XMM_Lambda}
            $\Lambda$CDM model. 
	    Each cluster is represented as a dot in this plot. Clusters 
	    marked with a big black dot have fluxes (in the $0.5-2$ keV band) 
	    above the limiting flux $S_X = 1.4 \times 10^{-14}$ 
	    erg s$^{-1}$ cm$^{-2}$ and they are the clusters expected to be 
	    detected by Newton-XMM in the XCS (Romer et al. 1999). 
	    Cluster marked with an open circle are those with a SZE flux 
	    above $30 $ mJy at $353$ GHz and they are the 
	    clusters expected to be detected by Planck. 
	    Solid lines represent the masses for which the 
	    flux equals the limiting flux of Planck ($30 $ mJy), 
	    and the XCS. 
	    Dotted lines show the maximum cluster mass expected in a solid angle 
	    equal to $2/3$ of the sky (top) and $(12.8^{\circ})^2)$ (bottom). 
	    Finally the heavy solid line indicates the selection function 
	    for an hypothetical optical survey made with a 10-m class telescope 
	    and two hours of observation (8 bands and 900s per band. See 
	    section \ref{section_Txitxo}).    
	   }
   \end{center}
\end{figure}

\begin{figure}
   \begin{center}
   \epsfxsize=9.cm 
   \begin{minipage}{\epsfxsize}\epsffile{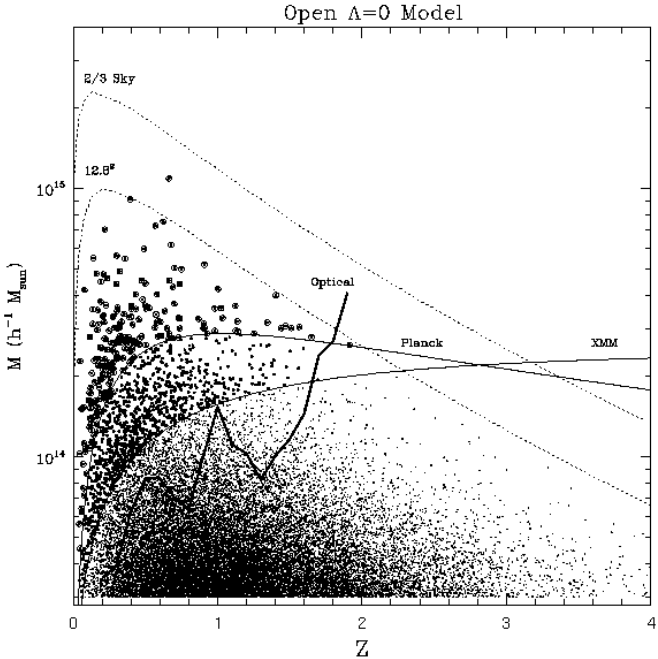}\end{minipage}
   \caption{\label{fig_Planck_XMM_Lambda0}
            OCDM case. Note the differences in the cluster distribution 
	    above the coasting phase redshift $z_c \approx 0.7$ (see text). 
	   }
   \end{center}
\end{figure}

\noindent
The total number of clusters is about $\approx 20000$ in one of these 
simulations. 
The mean Comptonization parameter is well below the FIRAS limit 
($\sim 10^{-6}$ compared with $2.5 \times 10^{-5}$, 
Fixsen et al. 1996). 
The resulting distribution of clusters is shown in 
figure \ref{fig_Planck_XMM_Lambda}.
The allowed range of simulated masses was 
$3.0\times 10^{13} h^{-1} M_{\odot} < M < 1.0\times 10^{16}  h^{-1} M_{\odot}$. 
The lower limit is at the frontier between a 
small cluster and a galaxy group and it is the mass corresponding 
to a cluster with temperature $T \sim 1$ keV which can be considered as 
the minimum temperature for a virialized cluster.\\  
In fig. \ref{fig_Planck_XMM_Lambda} each point corresponds to one cluster 
with mass $M$ and redshift $z$. 
This distribution is in agreement with the observational constraint given 
by the detection of at least three clusters with masses above 
$0.8\times 10^{15} M_{\odot}$ and $z > 0.5$ (Bahcall et al. 1998).\\
The most massive cluster is at redshift 0.66 with a mass $M = 1.09\times 10^{15} 
 h^{-1} M_{\odot}$ although this was an unusual realization. 
In a normal one of this size, the most massive cluster is usually 
well below $z \approx 0.5$. 
Clusters marked with an open circle correspond to those with a total flux 
above $30 $ mJy and according to our criterion, these clusters 
would be detected by Planck (see Appendix A). 
There are 185 clusters above this limit and only one of them 
would be observed above $z = 1$ in this simulation. 
As a comparison we show the same picture 
but corresponding to the OCDM model in table 1. 
(fig. \ref{fig_Planck_XMM_Lambda0}). In this case $\approx 15$ 
clusters are detected above redshift $z = 1$. 
This comparison demonstrates how a small region of 
the sky can show up the differences between the two models in table 1. 
Both models are nearly undistinguishable below $z \approx 0.7$ but 
they differ significantly above that redshift. 
We will come later to this point in section \ref{section_Optical}.\\

\noindent
As a comparison with the number of detected clusters expected with Planck,  
we show in figs. \ref{fig_Planck_XMM_Lambda} and   
\ref{fig_Planck_XMM_Lambda0} the clusters expected to be detected by 
Newton-XMM  
(big solid circles, $S_X > 1.5\times10^{-14}$ erg s$^{-1}$ cm$^{-2}$ 
in the $0.5-2$ keV band, Romer et al. 1999). \\
Solid lines represent the minimum mass as a function 
of redshift for which the corresponding flux is above the limiting 
flux in Planck and Newton-XMM respectively, that is, they are the selection 
functions for both missions. 

\noindent
Although Newton-XMM will see many more clusters at low redshift, 
however Planck will 
be more sensitive to those clusters at high redshift. The reason is that the 
corresponding selection functions are different in both cases.
The X-ray flux goes as $S_X \propto D_L^{-2}(z)$ 
and to detect clusters deeper in redshift they  must be more luminous 
(more massive) in order to compensate the monotonous increasing
function $D_L(z)$. 
On the contrary, the SZE flux goes like $S_{SZE} \propto D_a^{-2}(z)$ and 
at redshift $\approx 1$, the angular diameter distance reaches a maximum 
and after that it starts to drop to smaller values. Therefore the masses 
needed to provide a particular flux, $S_{SZE} = 30 $ mJy, can be smaller at 
redshift $z > 1$ than the required masses at $z \approx 1$. \\
The dotted lines in figs. (\ref{fig_Planck_XMM_Lambda}) 
and (\ref{fig_Planck_XMM_Lambda0}) indicate the 
maximum expected mass as a function of redshift in the simulation for 
the specific solid angle  $2/3$ of the sky (top) and 
$12.8^2 deg^2 $ (bottom). The point where the selection 
function cross these {\it maximum mass} lines corresponds 
to the maximum redshift expected in the sample. In both cases, Planck goes 
deeper than Newton-XMM. The difference can be even more emphasized if the 
limiting flux for Planck is below the one considered in this paper which 
was a very conservative one (see Appendix A).\\

\begin{figure}
   \begin{center}
   \epsfxsize=8.cm 
   \begin{minipage}{\epsfxsize}\epsffile{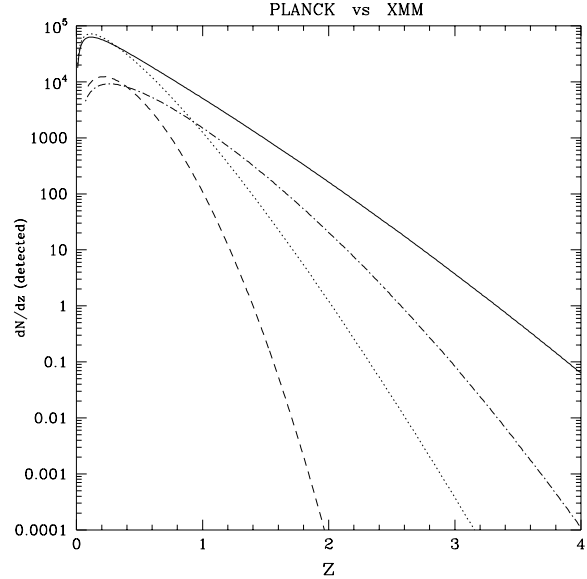}\end{minipage}
   \caption{\label{fig_Planck_vs_XMM}
            Expectations for Planck (solid line OCDM, dotted line $\Lambda$CDM) 
	    and the XCS (dot-dashed line OCDM, dashed line $\Lambda$CDM). 
	   }
   \end{center}
\end{figure}

\noindent
As noted in Romer et al. (1999), the XMM Cluster Survey (XCS) 
will cover $\sim 800$ square degree and will contain more than $8000$ clusters. 
On the contrary Planck will observe the full sky, that is the sky coverage 
will be $\approx 50$ times larger than that of the XCS. In order to 
compare Newton-XMM with Planck this difference in the sky coverage should be 
considered.
In fig. \ref{fig_Planck_vs_XMM} we show the number of detected clusters  
expected by  Planck and Newton-XMM (XCS) where we have taken into account the 
sky coverages and limiting fluxes expected for both missions. 

\noindent
Now, the differences between Planck and Newton-XMM are evident. 
The large sky coverage
of Planck together with the constancy of the SZE surface brightness 
with redshift, will allow this satellite to detect a much more 
significant number of clusters than Newton-XMM in the XCS. 
Also, we can conclude from this plot that, for 
the best fitting models in table 1, no clusters are expected to 
be detected above $z \approx 3$, neither for the OCDM nor the $\Lambda$CDM. 
However the cluster abundance between $z = 1$ and $z = 2$ 
will provide a definitive probe of the cosmological parameters and the 
cluster scaling relations.

\noindent
An important consequence of the previous plots 
(figs. \ref{fig_Planck_XMM_Lambda} and \ref{fig_Planck_XMM_Lambda0}) 
is that only a small portion of the 
whole sky would be needed to distinguish between the 
two models $\Lambda$CDM and the OCDM. This is an important point 
because only spectral identification of $\approx 200$ (not resolved) 
random selected clusters from the whole catalogue  would be needed.
However, it is important to answer the question, what is the minimum 
number of clusters needed to distinguish the models in table 1 at, 
for instance, the 3 $\sigma$ level? We will try to answer this 
question in the next section. 

\section{An optically-identified SZ-selected catalogue}\label{section_Optical}
\begin{figure}
   \begin{center}
   \epsfxsize=9.cm 
   \begin{minipage}{\epsfxsize}\epsffile{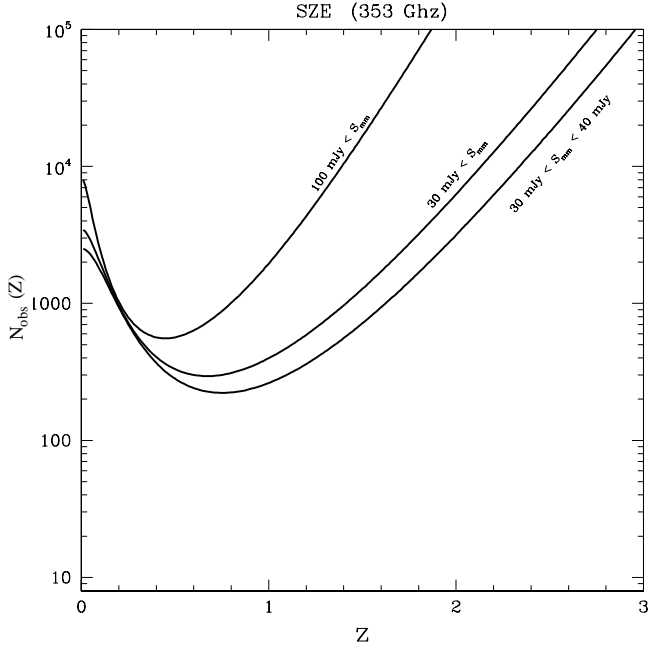}\end{minipage}
   \caption{\label{N_required}
	    Required number of clusters to be observed in order to distinguish 
	    the OCDM and $\Lambda$CDM models. }
   \end{center}
\end{figure}

As we have seen in fig. \ref{fig_NS_z_A}, the information provided by 
an hypothetical $N(S)$ curve (even if this curve corresponds to a full sky 
survey) will be insufficient to distinguish the two 
models considered in that figure. Redshift information 
will be needed in order to make the distinction. 
Different cosmological models predict different cluster populations as a 
function of redshift. If we analyze the evolution of the cluster population 
with $z$, then we should be able to discriminate among those models. 
However, to study the evolution of the cluster population we 
need spectral identification of the clusters (or at least optical observations 
in several bands in order to get photometric redshifts) since the SZE does
not provide any estimate of $z$. Performing these observations for an
hypothetical full sky SZ catalogue would be a huge task but if only a small 
number of unresolved clusters need to be identified then the work 
is significantly simplified. Now, we should ask the question, how small can 
our optically-identified sample be if we want to distinguish between, 
for instance, the two models in table \ref{table}?  
To answer that question we have compared the curves 
$N(S_{mm} > 30 $ mJy$,> z)$
for the two models in which we are interested. 
We require that at a given redshift, both curves must be distinguished 
at a $3\sigma$ level, that is, we require the condition 
$N^O -  3\sqrt{N^O} = N^{\Lambda} + 3\sqrt{N^{\Lambda}}$ 
(assuming Poissonian statistics), 
where $N^O$ and $N^{\Lambda}$ are the number of detected clusters 
above a given $z$ in the two cases OCDM and $\Lambda$CDM respectively.\\ 
Since we know $N^O$ and $N^{\Lambda}$ at each $z$, then we can compute 
the required total number of clusters which should be observed in order to 
satisfy the previous condition at each redshift. 
In fig \ref{N_required} we show this calculation 
for three different selection criteria of the clusters. In each one of the 
lines we show the total number of clusters randomly selected from the   
catalogue (with the only condition that the total flux must be 
$S_{mm} > 100 $ mJy top, $S_{mm} > 30 $ mJy middle and 
$30 $ mJy $< S_{mm} < 40 $ mJy bottom)  
which should be optically observed 
in order to distinguish (at a 3$\sigma$ level) $N^O$ and $N^{\Lambda}$  
at redshift $z$ (i.e. the total number of observed clusters needed to 
have a $3 \sigma$ difference in $N(>z)$ for the two models). 
As can be seen from the plot, randomly selecting about 
$300$ clusters with $S_{mm} > 30 $  mJy from the full catalogue 
and determining 
the redshift for each one, will allow to distinguish the two  
models at a 3$\sigma$ level above $z \approx 0.6$ just by looking at the 
$N(>z=0.6)$ curve.
The explanation for this fact is given by the different evolution of the 
cluster formation in both models. 
In the $\Lambda$CDM case there is a coasting phase 
(or inflection point in the acceleration parameter) at $z_c \approx 0.7$ 
that helps to form structure at this redshift. This phase is not 
present in the OCDM case where there is a redshift $z_c \approx 2$ 
below which the collapse of linear fluctuations is inhibited by the 
fast expansion of the Universe.\\
Choosing the selection criteria 
$30 $ mJy $< S_{mm} < 40 $ mJy, it is possible to reduce 
slightly the number of clusters to be identified. If on the contrary, 
only the brightest clusters with $S_{mm} > 100 $ mJy are 
identified optically, then we would need a significantly larger number of 
clusters in order to make the distinction between the models.
 
\subsection{Cluster optical detection simulations}\label{section_Txitxo}

Probably the most cost efficient way of identifying in the optical the galaxy 
clusters detected by Planck is using photometric redshifts. Even a rough
estimate of the photo-z allows to considerably reduce the background galaxy 
contamination and enhance the contrast density of the cluster. In addition, 
although the error in the photo-z of an individual galaxy is usually 
$\approx 0.1$ at $z<1$ (Ben\'\i tez 2000), the total error in the cluster 
redshift will be $0.1/\sqrt{N_{cl}}$ ($N_{cl}$ being the number of 
galaxies in the cluster).
To estimate the feasibility of detecting Planck SZ cluster candidates 
using optical imaging, we perform simulations based on empirical 
information. Since extensive data are only available for relatively 
low redshift clusters, this has the disadvantage of ignoring evolution 
effects. However, it has been shown that the evolution of the cluster 
early type galaxies is not dramatic up to $z\approx 1.3$ (Rosati 
et al. 1999). Therefore, at worst, this makes the results obtained 
here a conservative, lower limit on the detectability of high redshift 
clusters, since any reasonable luminosity evolution would tend to make 
the cluster galaxies bluer and brighter, increasing the number of 
detected galaxies with respect to the non-evolution case. \\
Wilson et al. (1997) represent the luminosity function 
of bright galaxy clusters as the superposition of a Schechter function 
and a Gaussian for the brightest cluster galaxies. The spectral fraction for 
the cluster galaxies can be derived  comparing the $V-I$ colors of 
galaxies in A370 and CL 1447+23 (Smail et al. 1997) with those 
expected for El, Sbc, Scd and Im spectral templates 
(Coleman, Wu \& Weedman 1980). With the above luminosity function and 
spectral fractions (extended to 8 magnitudes fainter than $M_*$) 
we generate a bright galaxy cluster at $z=0.18$. 
By redshifting this cluster, we generate a series 
of mock cluster catalogs containing the $I$ band magnitude and the 
spectral type $T$ from $z=0.2$ to $z=2.0$. The $I-$ band magnitudes are 
transformed using k-corrections derived from the Coleman, Wu and Weedman 
templates. To model the surface number counts distribution of the clusters, 
we use a $n(R)\propto R^{-0.3}$ law, found by Vilchez et al. 2001 to 
agree well with the projected galaxy density in the 
central regions of galaxy clusters.  To link the optical results with 
X-ray and SZ quantities, we integrate the luminosity function of the 
cluster in the $V-$band, and assume $M/L=300$, which leads to  
$M \sim 10^{15} M_{\odot}$ for a A1689-like cluster. \\

\noindent
We simulate the background galaxy distribution $n(z,T,I)$ using the 
Hubble Deep Fields (Williams et al. 1996, Williams et al. 2000). For the 
redshifts, we use the spectroscopic results of (Cohen et al. 2000) for 
the HDFN and photo$-z$ obtained using the BPZ code (Ben\'\i tez 2000) 
for the rest of the HDFN galaxies and those of the HDFS. \\
Once we have $I,z,T$ catalogs for all the galaxies contained in the field, 
we generate $UBVRIJHK$ magnitudes using the above mentioned template library, 
enlarged for the HDF galaxies using two starsburst templates from Kinney 
et al. 1996. Gaussian photometric errors are added to these {\it ideal} 
magnitudes using empirical relationships derived from real observations 
with 10m class telescopes, scaled by the square root of the exposure time 
needed to simulate a 900s per band observation.  

For cluster detection, we look for an overdensity of ellipticals with
respect to the expected background population. The reason to do this, 
instead of using the whole cluster population, is that the density 
contrast is much higher for this type of objects. Even a moderate cluster 
at $z\sim 1$ (Ben\'\i tez et al. 1999) conspicuously stands out against 
the relatively sparce numbers of field ellipticals 
(see also Gladders \& Yee 2000). 
Therefore, we estimate photometric redshifts for all the galaxies within an
angular aperture corresponding to $\approx 1h^{-1}$Mpc, classify them into 
different spectral types, and construct a redshift histogram for the early
types. 
The presence of a `spike' in the redshift histogram will indicate the 
existence of a cluster. The signal-to-noise of such detection is 
\[
S/N \approx [N(z)-<N(z)>]/\sigma_g(z)
\]
where $\sigma_g(z)$ is the expected fluctuation in the galaxy numbers 
within a redshift slice centered on $z$. Its value can be estimated 
as:
$$
 \sigma_g(z)^2=<N(z)>+\frac{<N(z)>^2}{(S^2)}\int w(\theta_{12})dS_1 dS_2
$$
$N(z)$ is the detected number of galaxies with an area $S$, 
$<N(z)>$ is the expected average density and $w(\theta_{12})$ is 
the two-point correlation function within the redshift slice. 
For most of the redshifts considered, $1$Mpc$h^{-1}$ corresponds to 
$\approx 4$ arcmin. Taking the amplitude of $w(\theta_{12})$ 
to be $A\approx 7.6\times10^{-3}$ within a $z=0.2$ slice (Brunner, 
Szalay \& Connolly 2000), which is approximately the same redshift 
interval used here to detect the clusters, the value of $\sigma(z)$ 
is roughly $\sigma_g(z)^2\approx <N(z)>(1+3.27\times 10^{-3}<N(z)>)$. 
This number may be an understimate since the clustering strength of the 
early types is known to be higher than for the general galaxy population.
The numbers below are based on a $3\sigma$ detection limit as defined by 
the above equation. There are plenty of other methods, parametric and 
non-parametric, which will probably be more efficient in finding 
clusters (see e.g. Nichol et al. 2000). 
But again, we think that using a relative simple 
procedure provides a good idea about the practicality of this approach.
A reasonable observing strategy will be to start with those clusters not 
detected in shallower imaging surveys, e.g. the SDSS catalog, and depending on 
the redshift/luminosity range of interest (e.g. low mass/low redshift clusters
or 
high mass/high redshift ones), use only a small subsample of the $BVRIJHK$
filter set mentioned above, which brackets the $4000 \AA$ break at the 
required redshift, and which would be enough to detect the early types. 
If one desires to reach a higher precision in the photo-z estimation, 
or wants to reach the limits shown in Figs \ref{fig_Planck_XMM_Lambda} and 
\ref{fig_Planck_XMM_Lambda0}  at all the redshift intervals, 
then the whole filter set should be used. 
The optical selection function in Figs \ref{fig_Planck_XMM_Lambda} and 
\ref{fig_Planck_XMM_Lambda0} presents quite a jagged look. Apart
from relatively smooth effects like the cosmological dimming or the 
K-corrections which determine the general trend, the detectability of 
the clusters is significantly affected, at least at $z<1$, by the relative 
placement of the redshifted $4000\AA$ break with respect to the filter 
set, specially to the $R$ and $I$ band filters,
the ones which go deeper for a fixed exposure time. When the break falls
almost exactly in between these two filters, the photo-z precision is 
improved, whereas if the break is close to the central position of a filter 
the redshift error increases and the estimation is more easily affected 
by color/redshift degeneracies. Therefore, the relative exposure times 
and the filter choice should be optimized depending on the redshift 
interval that it is being targeted.

\section{Estimating the cosmological model from $N(S)$ and $N(z)$}\label{section_Constraining}
\noindent 
Previous works (Kitayama \& Suto 1997, Eke et al. 1998, 
Mathiesen \& Evrard 1998, Borgani et al. 1999, Diego et al. 2000) 
have shown the power of X-ray surveys to constrain the cosmological model. 
From the previous sections we conclude that also the SZE data can be used 
with the same purpose (see also Markevitch et al. 1994, 
Barbosa et al. 1996, Aghanim et al. 1997, Holder \& Carlstrom 1999, 
Majumdar \& Subrahmanyan 2000, Fan \& Chiueh 2000). \\
As we have seen in the previous sections, both the $N(S)$ and the 
$N(> 30 $ mJy$, z)$ 
curves can be used to study the cluster population, being $N(S)$ the curve 
having larger number of clusters (although with no $z$ information) and 
$N(> 30 $ mJy$, z)$ the curve which is 
more sensitive to the evolution of 
the cluster population. Following Diego et al. (2000) we will combine both 
curves to reduce the degeneracy. Some models which are compatible with the first 
curve will be incompatible with the second one and vice versa. Thus, those 
models will be rejected. \\
Since, this kind of data is not available yet, we will check the method with 
two simulated data sets following the characteristics of the Planck satellite 
(section \ref{section_probe}) for $N(S)$. For the second curve we will suppose 
that a randomly selected subsample of 300 clusters from the whole Planck 
catalogue have been observed optically and that we know the redshift 
for each cluster in this subsample (see section \ref{section_Optical}).
The input model used to simulate both data sets was the $\Lambda$CDM model 
in table 1.
That model is compatible with the mass function given in 
Bahcall \& Cen (1993), the temperature function of 
Henry \& Arnaud (1991), the luminosity function of Ebeling et al. (1997), 
and the flux function of Rosati et al. (1998) and De Grandi et al. (1999),  
as it was shown in Diego et al. (2000).\\

\noindent
We have compared both simulated data sets with $\approx 2$ millions 
different flat $\Lambda$CDM models where the six free parameters of our model 
have been changed on a regular grid. 
The first three parameters correspond to the cosmological ones ($\sigma_8$, 
$\Omega$, and  $\Gamma$) which control the cluster population in the 
Press-Schechter formalism. The other three parameters correspond to the 
$T-M$ relation (eq. \ref{Tx}) whose free parameters will enter 
in the fitting procedure at the same level as the cosmological ones. 
This relation is needed to build the total flux from the mass of the 
cluster (eq. \ref{S_total2}). 
By considering the $T-M$  as a free parameter relation, we will check 
the influence of the uncertainty in the scaling relation on the 
determination of the cosmological parameters.\\

\begin{figure}
   \begin{center}
   \epsfxsize=9.cm 
   \begin{minipage}{\epsfxsize}\epsffile{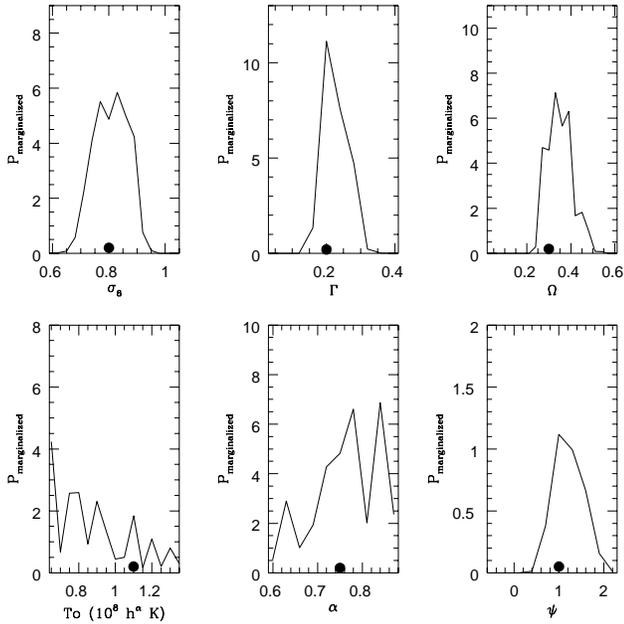}\end{minipage}
   \caption{\label{figure_Margin}
	    Marginalized probability in each parameter. The black dot represents 
	    the input model ($\Lambda$CDM model in table 1).  }
   \end{center}
\end{figure}

\noindent
In fig. \ref{figure_Margin} we show the results of our fit. We have 
marginalized the probability over each one of the six free parameters.
The probability was defined as in Diego et al. (2000) using the 
Bayesian estimator given in Lahav et al. (1999).
\begin{equation}
 -2 ln P_L = \chi_{L}^2 ,
 \label{Lahav}  
\end{equation}
\noindent
where, 
\begin{equation}
 \chi_{L}^2 = \sum_i^2 N_i ln(\chi_i^2) . 
 \label{Lahav2}
\end{equation}
$\chi_i^2$ is the ordinary $\chi^2$ 
for each data set and $N_i$ represents the number of data points for 
the data set $i$. Based on a Bayesian approach with the choice 
of non-informative uniform priors on the log,  
those authors have seen that this estimator is appropriate for the case when 
different data sets are combined together, as it is our case. The factor $N_i$ 
plays the role of a weighting factor. Larger data sets are considered 
more reliable for the parameter determination.\\
As shown in that figure, the estimate of the cosmological 
parameters is unbiased (compare with the input model, black dots). 
They are also very well constrained with a small degeneracy between the 
parameters. This result shows how with future SZE data it will be possible 
to discriminate among several scenarios of cluster formation.  
The situation is different in the parameters of the $T-M$ relation. 
In this case we do not get any spectacular result. Only the $\psi$ parameter 
has been well located. There is a degeneracy between the amplitude $T_0$ and 
the exponent $\alpha$ which will be discussed in the next section.

\begin{figure}
   \begin{center}
   \epsfxsize=9.cm 
   \begin{minipage}{\epsfxsize}\epsffile{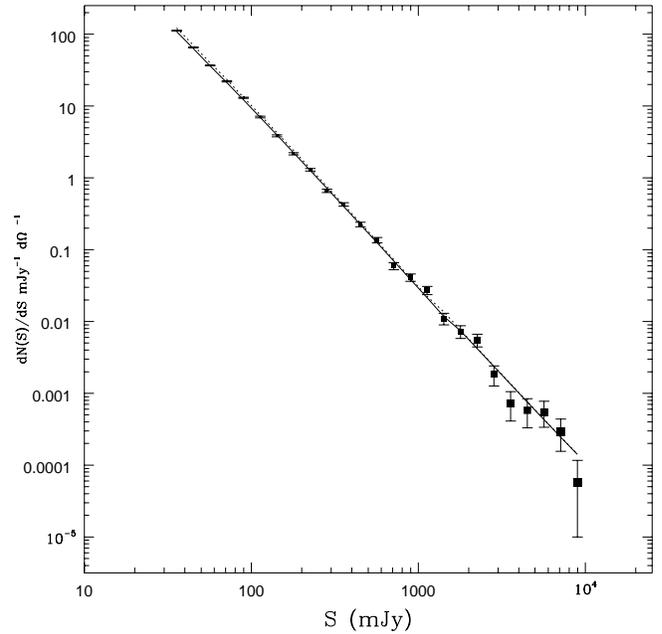}\end{minipage}
   \caption{\label{figure_NS_Planck}
	    $dN/dS$ curve for the $\Lambda$CDM (solid) and OCDM (dotted) models in table 1. 
	    Data points represent a Monte Carlo realization of mock data for the $\Lambda$CDM 
	    model. This curve includes all the Planck detected clusters (2/3 of the sky).  }
   \end{center}
\end{figure}

\begin{figure}
   \begin{center}
   \epsfxsize=9.cm 
   \begin{minipage}{\epsfxsize}\epsffile{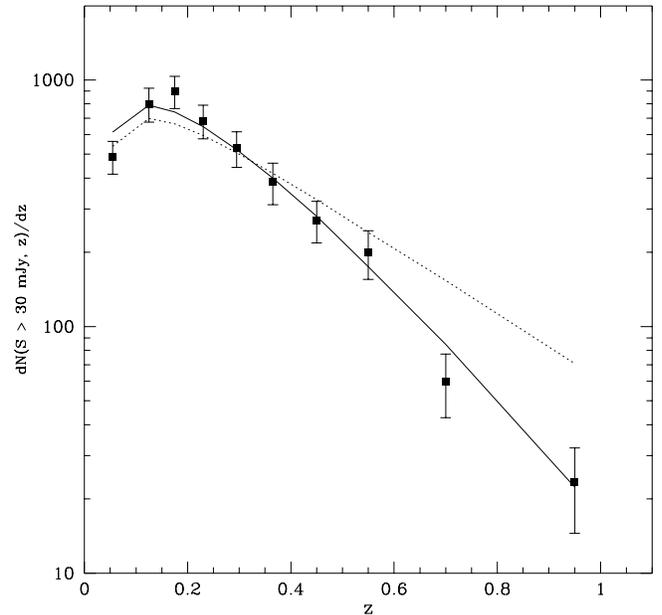}\end{minipage}
   \caption{\label{figure_NGTS_z}
	    $dN(S > 30  mJy, z)/dz$ (353 GHz) curve for the $\Lambda$CDM 
	    (solid) and OCDM (dotted) 
	    models in table 1. As in figure \ref{figure_NS_Planck} the data was simulated 
	    assuming a  $\Lambda$CDM model. The simulation was over a solid angle  
	    $d\Omega = 0.6675 \%$ of the sky where a total of 
	    $\approx 300$ clusters were found with fluxes greater 
	    than $30 $ mJy. }
   \end{center}
\end{figure}

\noindent
In fig. \ref{figure_NS_Planck} and fig. \ref{figure_NGTS_z} we present 
the simulated data sets and the two {\it undistinguishable} models 
given in table 1. From the first figure it is evident that 
both models would remain undistinguishable if only that data set is used in 
the fit but in the second figure we can see that it is possible to distinguish 
the two models at a high significant level due to their different evolution in 
redshift.
 
\section{Discussion \& Conclusions}\label{section_Discussion}
In previous sections we have seen that the SZE will be a very powerful tool 
to study the cluster population at different redshifts.
Up to now, no cluster has been detected above $z \approx 2.0$. 
Previous X-ray surveys have been limitted in redshift and current 
experiments (CHANDRA, Newton-XMM) are not expected to detect clusters 
much above that. Only through the SZE we can have the 
possibility to observe clusters above that redshift (or maybe to conclude that 
no cluster has been formed above that redshift). These high redshift 
clusters are fundamental to understand the physics of cluster formation 
and also to establish the evolution of the cluster scalings such as 
the $T-M$ or $L_x-M$.\\
We have seen that Planck will be able to detect distant clusters which will 
provide very useful information about the cluster population and the 
underlying cosmology. \\
However, we have seen that with only the $dN(S)/dS$ curve, it will be difficult 
to discriminate among models which were previously undistinguishable. 
To distinguish them, we need redshift information. 
We have seen that for the whole SZE sky catalogue, only a relatively  
small number ($\approx 300$) of optically observed clusters randomly selected 
from the whole Planck catalogue is needed 
in order to discriminate between the $\Lambda$CDM and OCDM models just 
by looking at the different cluster population as a function of redshift. 
If we want to discriminate among the $\Lambda$CDM models, we have shown that 
by combining the statistically large data set 
$dN(S)/dS$ with the cosmological sensitivity of $dN(>30  mJy, z)/dz$ it 
is possible to reduce significantly the degeneracy in the cosmological 
parameters as can be seen in fig. (\ref{figure_Margin}). \\

\noindent
One important conclusion is that this result is almost independent 
of the assumed $T-M$ relation.  
In fact our method is practically insensitive to the $T_0$ amplitude and 
$\alpha$ exponent in eq. (\ref{Tx}). 
We have marginalized the probability assuming different fixed values for 
$T_0$ and $\alpha$. The resulting marginalized probabilities were very 
similar in all the cases considered showing the small dependence on the 
assumed values of $T_0$ and $\alpha$. 
The almost null dependence on $\alpha$ can be 
easily understood by looking at eq.(\ref{S_total2}). 
In the computation of the temperature function (see eq. \ref{dNT}) 
the derivative $dM/dT$ was inversely proportional to $\alpha M^{\alpha-1}$.  
The X-ray derived functions (like the temperature function) are sensitive 
to the $\alpha$ exponent through the previous derivative. 
On the contrary, the flux function, $d^2N/dSdz$ is inversely proportional to 
the derivative  $dS/dM \propto (1 + \alpha) M^{\alpha}$. 
Therefore, a change of 0.1 units in $\alpha$, represents 
a change in the $dM/dT$ derivative of $14 \%$ while in the flux function 
the same change in $\alpha$ implies a variation of only $6 \%$  in the 
derivative $dS/dM$, both percentages assuming 
$M = 1\times10^{15} h^{-1} M_{\odot}$. 
This explains why the SZ flux function is less sensitive to $\alpha$ 
than the X-ray derived functions.
The uncertainty in $T_0$ is a bit more difficult to understand. 
From eq. (\ref{S_total2}) the total flux is directly proportional 
to $T_0$ and therefore we should expect some dependency of our fit 
on this parameter. However, if a change in $T_0$ is compensated by 
a change in $\alpha$ then we would have 
a degeneracy on these two parameters 
($S \propto T_0 M^{1+\alpha}(1 + z)^{\psi}/D_a(z)^2$). 
In fact from fig. \ref{figure_Margin} we can see that those models 
with a low value for $T_0$ and a high value of $\alpha$ are 
slightly favored indicating this fact that there is some kind of 
compensation between these two parameters. \\
In order to break the degeneracy in $T_0 - \alpha$ we should include in 
our fit information concerning the mass of the clusters just to make the 
fit sensitive to an independent change in $T_0$ and/or $\alpha$ 
and not only to a change in the quantity $T_0 M^{1+\alpha}$. The 
previous situation was considered in Diego et al. (2001) where we included in 
the fit the cluster mass function. In that case we found in fact that there 
was not any degeneracy in those parameters. \\
The third parameter of the $T-M$ relation ($\psi$) seems to be, however, 
very relevant for our fit. This is not surprising as we are using 
one data set which is expressed as a function of redshift 
(fig. \ref{figure_NGTS_z}). While both, $T_0$ and $\alpha$ can be 
mutually compensated, the effect of changing $\psi$ on the simulated 
data sets (figs. \ref{figure_NS_Planck} and \ref{figure_NGTS_z}) can 
only be compensated with a change in some of the cosmological 
parameters (through their effect on the cluster population and in $D_a(z)$) 
but as the allowed range of variation of the cosmological parameters 
is small (see fig. \ref{figure_Margin}), consequently the confidence 
interval for $\psi$ will be small as well. \\

\noindent
In this work, the $T-M$ relation  was considered as a free relation just 
for consistency with our previous work. When fitting SZ data, we have shown 
that the choice of one specific value for $T_0$ and $\alpha$ in the  $T-M$ 
relation is not quite relevant, although it is important to include in 
the fit the possible dependency of this relation with $z$. This situation 
is opposite to the one in Diego et al. (2000) where the redshift 
dependence was not relevant (since most of the data was at low redshift) 
but the choice of $T_0$ and $\alpha$ was important to obtain a good fit 
to the X-ray and optical data considered in that work.
The specific form of the $T-M$ relation will be more important in the 
case of fitting future X-ray data. Newton-XMM will provide very 
relevant information, specially at low and medium redshift, about the cluster 
population and the scaling relations $T-M$ and $L_x - T$. 
However, we have seen that the expected number of detected clusters and 
the redshift coverage will be smaller for this mission compared with Planck and 
therefore Planck will provide several key informations to understand   
cluster formation and evolution. For instance, as we have 
already seen, the information about the $T-M$ relation can be complemented 
with studies of the SZE on clusters. Meanwhile $T_0$ and $\alpha$ can 
be determined through the study of low redshift X-ray data, $\psi$ could be 
constrained with the high redshift SZE data.
The best results will come, therefore, from the combination of data 
from X-ray and mm missions (see eg. Haiman, Mohr \& Holder, 2000).
With Newton-XMM we can obtain a good sampling of the cluster population at 
low-intermediate redshift with their corresponding temperatures and X-ray 
fluxes (also detecting the low mass population) and with Planck we will 
explore the cluster population further in redshift. \\
A very interesting possibility has been analyzed by Xue \& Wu (2001). 
The authors suggested the use of the X-ray luminosity function as a starting 
point to derive the mm (SZE) flux function. In the process, several 
assumptions about the $L_x-T$ and $T-M$ relations need to be done. 
These assumptions could be tested with future SZE data opening the possibility  
to study those relations at redshifts where no clusters can be observed 
in the X-ray band.\\

\noindent
Although this paper has concentrated on the possibilities of the 
future Planck SZE catalogue, proposed and undergoing mm experiments 
will measure the SZE for hundreds of clusters before Planck is launched. 
These experiments will open a new era in which many works will be done 
based on those exciting data.\\

\section{ Acknowledgments}
This work has been supported by the 
Spanish DGESIC Project  
PB98-0531-C02-01, FEDER Project 1FD97-1769-C04-01, the 
EU project INTAS-OPEN-97-1192, and the RTN of the EU project   
HPRN-CT-2000-00124. \\
JMD acknowledges support from a Spanish MEC fellowship FP96-20194004 
and financial support from the University of Cantabria.




\newpage


\appendix

\section{Planck flux limit}
In this appendix we want to justify that Planck will be able to detect 
those clusters with an integrated total flux above $\approx 30 $ 
mJy.\\

\noindent 
Obviously, the number of detected clusters will depend on the technique 
used to detect them. We will focus our attention on the 
Maximum Entropy Method (MEM) (Hobson et al. 1998, Hobson et al. 1999) 
where the authors have shown that with such method they obtain a good 
recovery of the thermal SZE. 
In that paper it is shown that the rms residuals per $4.5$ arcmin 
FWHM Gaussian beam for the MEM reconstruction is $\approx 6 \mu K$ 
per pixel. \\
Now we compute the flux (in mJy) corresponding to that rms 
temperature and therefore it should be considered as a flux {\it per pixel}.

\noindent
The flux is defined as the integral of the specific intensity on the 
solid angle
\begin{equation}
 S_{SZE}(\nu) = \int \Delta I_{SZE}(\nu) d\Omega \ .
 \label{S_pixel}
\end{equation} 
When we compute the total flux of the cluster the solid angle 
is that subtended by the cluster (see eq. (\ref{S_total}) in section 
\ref{section_SZE}). 
In eq. (\ref{S_pixel}), $\Delta I( \nu )$  can be related 
with $\Delta T/T$ as  $\Delta I( \nu ) = I_0g(x)(\Delta T/T)$ where $I_0$ is a 
constant given below, $g(x)$ is the spectral shape factor, 
$\Delta T = 6.0  \  \mu K$ is 
the temperature we want to transform into a flux and $T = 2.73$ K,
\begin{equation}
I_0 = \frac{(k_B T_{CMB})^3}{(h_p c)^2} = 2.7 \times 10^{11} \frac{mJy}{str} 
\end{equation}
And the spectral dependence $g(x)$ is given by:
\begin{equation} 
g(x) = \frac{x^4e^x}{(e^x - 1)^2}
\end{equation}
where $x= h\nu/(k_B T_{CMB}) \approx \nu (GHz)/56.8$ is the adimensional 
frequency. Assembling terms in eq. (\ref{S_pixel}) we get the 
flux {\it per pixel} ($d\Omega =$ ($1.5$ arcmin)$^2 = 1.9 \times 10^{-7}$ str):
\begin{eqnarray}
S_{SZE}(\nu) &=& 2.7\times10^{11}\frac{\Delta T}{T}g(x)d\Omega \frac{mJy}{str} \\ \nonumber
             &=& 0.45 \frac{mJy}{pix}  \  \ (300 GHz)
\end{eqnarray}
This number has been calculated for the frequency $\nu = 300$ GHz. In the paper 
we presented our calculations for 353 GHz. The flux at this frequency is 
a factor $\approx 1.3$ times higher than at 300 GHz. Therefore 
\begin{eqnarray}
S_{SZE}= 0.58 \frac{mJy}{pix}  \  \ (353 GHz)
\end{eqnarray}
This flux should be considered as the rms in the residual map 
when the SZE is recovered by MEM, that is the noise per pixel. 
We will consider a conservative limit for the detection of a cluster of 
signal to noise ratio $ \geq 7.5 \sigma$ on the FWHM. 
If we consider our antenna as a Gaussian beam and the cluster is not resolved 
(this will happen in most of the cases in Planck) then we can consider that 
the cluster profile follows a Gaussian pattern:
\begin{equation}
 Signal = S = A e^{-r^2/2\sigma^2}
 \label{Signal}  
\end{equation}
Inside the FWHM the signal is just the integral of the previous equation 
from $r=0$ to $r=FWHM/2$ which can be solved easily (by doing the variable 
change $x = r^2/2\sigma^2$):
\begin{equation}
S(FWHM)= A\int_0^{FWHM/2} 2\pi r dr e^{-r^2/2\sigma^2} = 6.3\times A
\label{Signal_FWHM}
\end{equation}
Inside the FWHM the noise associated to MEM goes like:
\begin{equation}
N(FWHM) = N_{rms}\sqrt{N^{pix}_{FWHM}}
\label{Noise_FWHM}
\end{equation}
where  $N_{rms}$ is the noise per pixel found previously and 
$N^{pix}_{FWHM}$ is the number of pixels corresponding to the 
area enclosed by the FWHM, $N^{pix}_{FWHM} = FWHM^2 = 11.1$ (FWHM = 5 arcmin 
for the channel at 353 GHz).\\
Now if we require $S(FWHM)/N(FWHM)= 7.5$ then by dividing eqs.  
(\ref{Signal_FWHM}) by (\ref{Noise_FWHM})
we get that $A$ must be $A \approx 4 \times N_{rms}$, that is, 
an unresolved cluster with a signal following eq. (\ref{Signal}) should 
have an amplitude $A \approx 4  \times N_{rms}$ in order to have 
$S/N = 7.5$ inside the FWHM.\\
The total flux inside the full antenna beam of such a signal is 
$S_{Total} = 12.6 \times A \approx 50 \times N_{rms}$. If now we substitute 
$N_{rms} = 0.58 mJy$ per pixel, then we finally 
obtain, $S_{Total} \approx  29 mJy$ which approximates the value of 
30 mJy used in the paper.



\bsp
\label{lastpage}
\end{document}